\documentclass[12pt]{article}
\usepackage{latexsym}
\usepackage{amsfonts}
\usepackage{mathrsfs}
\usepackage[cp1250]{inputenc}

\usepackage{graphicx}

\title{Conformal mechanics}
\author{Joanna Gonera\thanks{jgonera@uni.lodz.pl},\\ 
Department of Theoretical Physics and Computer Science \\
University of {\L}\'od\'z \\
Pomorska 149/153, 90 - 236 {\L}\'od\'z/Poland.}
\date{}

\begin{document}
\maketitle
\begin{abstract}
The $SL(2,\mathbb{R})$ invariant Hamiltonian systems are discussed within the framework of the orbit method.
It is shown that both dynamics and symmetry transformations are globally well-defined on phase space. 
The flexibility in the choice of time variable and Hamiltonian function described in the paper by de Alfaro et al.
(Nuovo Cim. 34A (1976),569) is related to the nontrivial global structure of $1+0$-dimensional space-time.
The operational definition of time is discussed.
\end{abstract}

\newpage
\section{Introducion}

Conformally invariant mechanics was introduced in Refs.\cite{c1}, \cite{c2}. The role of conformal group in $1+0$-dimensional 
space-time is played by the $SL(2,\mathbb{R})$ group. Its three generators describe time translation, dilatation and
special conformal transformation. The detailed discussion of the conformal mechanics was presented in Ref.\cite{c2}
where its main properties were carefully elaborated including the form of conformal transformations, flexibility
in the choice of time variable etc. Since then there appeared a number of papers devoted to this topic \cite{c3}. In particular the quantization
of clasical systems based on $SL(2,\mathbb{R})$ symmetry was discussed in Ref. \cite{3}.

The main restriction imposed in Ref.\cite{c2} is that the potential entering conformally invariant Hamiltonian must be
repulsive. It is dictated by the fact that only then the quantum mechanical version is well-defined. This restriction has
its classical counterpart: in the attractive case the "falling on the center" phenomenon appears: 
the particle reaches the boundary of phase space at finite time.

The conformal mechanics can be studied in the framework of the so-called orbit method \cite{c4}. Its main advantage is that,
once the symmetry group (acting transitively on the phase manifold) is specified, all invariant Hamiltonian systems may
be classified using group-theoretical methods. When applied to $SL(2,\mathbb{R})$ group the orbit method allows to find all $1+0$-dimensional  
conformally invariant Hamiltonian systems \cite{c5}. It appears that the dynamics is perfectly regular in all cases
while the spurious falling at the center phenomenon results from the choice of local Darboux coordinates on topologically nontrivial
phase space.

In the present paper we continue this analysis. First, we show that the global conformal transformations are smooth; all singularities
result from the choice of local Darboux coordinates. We discuss then the alternative form of conformal transformations 
where the time variable is also affected. The transformation rule for time is singular but this can be accounted for
by interpreting time as homogeneous coordinate in projective space. The change of coordinate in the latter corresponds to the 
flexibility in the choice of time variable described in Ref.\cite{c2}.

Finally, we discuss the operational definition of time which agrees with that obtained within the standard method of constructing 
the conformal action on space-time by projecting from higher-dimensional cone \cite{c6}.

The paper is organized as follows. In Sec.II we summarize the basic properties of conformal mechanics in standard formulation.
Sec.III is devoted to the study of the realization of $SL(2,\mathbb{R})$ as the group of conformal transformations on $1+0$-dimensional 
space-time. To this end the language of nonlinear realizations \cite{c7} is basically used but the global properties of group 
action are carefully taken into account. In Sec.IV we remind the basic findings of Ref.\cite{c5} concerning the classification of 
the Hamiltonian systems on which $SL(2,\mathbb{R})$ acts transitively. The action of conformal group on phase space is studied in some 
details in Sec.V. It is shown that the conformal transformations act smoothly on phase manifold; the only singularities are
related to the choice of Darboux coordinates. The operational definition of time is discussed in Sec.VI. Finally, Sec.VII is devoted to brief conclusions.

For simplicity all quantities used are dimensionless.

\section{Conformal mechanics.}

Let us start by reviewing the most essential findings of Ref.\cite{c2} concerning classical conformal mechanics. One start with Lagrangian
\\
\begin{equation}
\label{lagr1}
L=\frac{m}{2}\dot q ^2 - \frac{g}{2q^2};
\end{equation}
\\
here $0 < q < \infty $ and $g$ denotes the coupling constant which, in principle, can take any real value. 
The authors of Ref.\cite{c2} restrict their consideration to the repulsive case, $g >0$. In fact, the structure of the system
in a attractive case is more involved as it will be explained below. 

The dynamics described by the Lagrangian (\ref{lagr1}) admits $SL(2,\mathbb{R})$ group as the symmetry group. Indeed, for any 
$ A\in SL(2,\mathbb{R})$
\\
\begin{equation}
A = \left(
\begin{array}{cc}
     \alpha & \beta \\
     \gamma  & \delta 
     \end{array}
     \right)
     ,
\qquad \alpha \delta  - \beta \gamma = 1
\label{2}
\end{equation}
\\
we define
\\
\begin{eqnarray}
t' = \frac{\alpha t + \beta}{\gamma t + \delta } \nonumber
\\
q'(t') = \frac{q(t)}{\gamma t + \delta }; 
\label{3}
\end{eqnarray}
\\
then
\begin{equation}
\label{4}
L\left( q'(t'),\frac{q'(t')}{dt'}\right) \frac{dt'}{dt} = 
L\left( q(t),\frac{dq(t)}{dt} \right) -
\frac{d}{dt} \left(\frac{m\gamma q^2}{2(\gamma t +\delta }\right).
\end{equation}
\\
There are three types of basic transformations described by $SL(2,\mathbb{R})$ symmetry:\\
- time translations: $\alpha =\delta =1$, $\gamma _0 = 0$, $\beta = t_0$ 
\\
\begin{eqnarray}
t'&=&t+t_0\nonumber\\
q'(t)&=&q(t)
\label{5}
\end{eqnarray}
\\
- dilatations: $\alpha = e^{\lambda }$, $\delta =e^{-\lambda }$, $\beta =\gamma = 0$,
\\
\begin{eqnarray}
t' &=& e^{2\lambda }t\nonumber\\
q'(t')&=&e^\lambda  q(t)
\label{6}
\end{eqnarray}
\\
- conformal transformations: $\alpha = 1$, $\delta = 1$, $\beta = 0$, $\gamma = -c$
\\
\begin{eqnarray}
t'&=&\frac{t}{1 - ct}\nonumber\\
q'(t')&=&\frac{q(t)}{1 - ct}.
\label{7}
\end{eqnarray}

Denoting by $H$, $D$ and $K$ the relevant generators of the Lie algebra $sL(2,\mathbb{R})$ we find
\\
\begin{equation}
\label{8}
\left[ D , H \right] = -i H, \qquad \left[ D , K\right] = i K, \qquad \left[ K , H \right] = -2i D.  
\end{equation}
\\

Noether theorem yields the corresponding conserved charges:
\\
\begin{eqnarray}
\tilde H&=&\frac{m}{2} \dot q^2 + \frac{g}{2q^2}\nonumber\\
\tilde D&=&-\frac{m}{2} q \dot q + t\left( \frac{m}{2}\dot q^2 +\frac{g}{2q^2}\right)
\label{9}\\
\tilde K&=&\frac{m}{2}\left(\dot q t - q\right)^2 + \frac{g}{2q^2} t^2\nonumber
\end{eqnarray}
\\
or, in terms of canonical variables:
\\
\begin{eqnarray}
\tilde H&=&\frac{p^2}{2m}+ \frac{g}{2q^2}\nonumber\\
\tilde D&=&-\frac{qp}{2}  + t\left( \frac{p^2}{2m} +\frac{g}{2q^2}\right)
\label{10}\\
\tilde K&=&\frac{m}{2}\left(\frac{p}{m} t - q\right)^2 + \frac{g}{2q^2} t^2.\nonumber
\end{eqnarray}

They are constants of motion and obey the Poisson counterpart of the algebra (\ref{8}).
\\
\begin{equation}
\label{11}
\left\{ \tilde D , \tilde H \right\} = 
-\tilde H, \qquad \left\{ \tilde D , \tilde K\right\} = \tilde K, \qquad \left\{ \tilde K , \tilde H \right\} = -2\tilde D.  
\end{equation}

It should be noted that the three integrals $\tilde H$, $\tilde D$ and $\tilde K$ cannot be functionally
independent. In fact, our system has one degree of freedom so there are two independent integrals.
The relevant functional relation between our three generators is obtained by recalling the form of the 
Casimir operator for $sL(2,\mathbb{R})$ algebra. One finds (on the classical level no ordering is necessary)
\\
\begin{equation}
\label{12}
\tilde H \tilde K - {\tilde D}^2 = \frac{mg}{4}.
\end{equation}

It is easy to verify explicitly that $\tilde H$, $\tilde D$ and $\tilde K$ are generators (in the sense of
the theory of canonical transformations). To this end we solve first
\\
\begin{equation}
\label{13}
\frac{dX(c)}{dc} = \left\{ \tilde K(t),X(c) \right\}, \qquad X = q^2,\, K = pq
\end{equation}
\\
which yelds
\\
\begin{eqnarray}
\label{14}
q^2(t,c)&=&q^2(t,0)\left( 1 + ct\right)^2 - \frac{2}{m} q(t,0)p(t,0)ct^2\left( 1+ct\right) + \frac{2c^2t^4}{m}H\\
q(t,c)p(t,c)&=&q^2(t,0)mc\left( 1 + ct\right) + q(t,0)p(t,0)\left( 1-2c^2t^2\right) + 2ct^2\left( ct - 1\right) H.\nonumber 
\end{eqnarray}
\\

In the repulsive case, $g>0$, the above transformation is perfectly regular. In fact, the first eq. (\ref{14}) may be rewritten 
in the form
\\
\begin{equation}
\label{15}
q^2(t,c) = \left( q(t,0)\left( 1 +ct \right) - \frac{p(t,0)}{m}ct^2 \right)^2 + \frac{gc^2t^4}{mq^2(t,0)}.
\end{equation}
\\
We see that the eqs.(\ref{14}) define the smooth action of conformal transformations on phase manifold defined
by inequalities $0 < q < \infty $, $ -\infty < p < \infty $. This conclusion can be easily extended to the whole 
conformal group which acts smoothly and preserves the symplectic structure. On the other hand the initial conformal
transformations (\ref{3}) are singular so the question arises how these singularities enter the game. The answer to this question 
is quite simple. Eqs. (\ref{14}) show that the canonical momenta enter the symmetry transformations in highly nonlinear
way. Therefore, these transformations cannot be interpreted as resulting from point transformations unless the latter
comprise also the transformation of time variable; in such a case the complicated momentum dependence can be accounted
for the necessity of shifting the time variable. More precisely, using this idea one can define the action of conformal group 
as a group of point transformations which on the "mass shell" coincide with the canonical transformations (\ref{14}). In fact,
defining the new time variable by
\\
\begin{equation}
\label{16}
t' = \frac{t}{1 - ct}
\end{equation}
\\
and using the explicit form of the solutions to the equations of motion
\\
\begin{eqnarray}
q^2(t)&=&q^2(t_0) + \frac{2}{M}(t - t_0)q(t_0)p(t_0) + \frac{2(t-t_0)^2}{m}H\nonumber\\
q(t)p(t)&=&q(t_0)p(t_0) + 2(t-t_0)H
\label{17} 
\end{eqnarray}
\\
one easily find that the eqs.(\ref{14}) can be rewritten in the form
\\
\begin{eqnarray}
\label{18}
q^2(t',c)&=&\frac{q^2(t,0)}{(1-ct)^2}\\
q(t',c)p(t',c)&=&q(t,0)p(t,0) + \frac{mcq^2(t,0)}{1 - ct}\nonumber
\end{eqnarray}
or
\begin{eqnarray}
\label{19}
q(t',c)&=&\frac{q(t,0)}{(1-ct)}\\
p(t',c)&=&p(t,0)(1 - ct) + mcq(t,0)\nonumber
\end{eqnarray}
\\
which agree with the transformation properties implied by eqs. (\ref{3}) and (\ref{4}).
Again this reasoning can be extended to the whole conformal group.

We see that the singularities in the action of conformal group appear (at least in the repulsive case)
only if one insists on representing it as the group of point transformations. They are rooted
in the singular action of the conformal group on time variable. This action will be discussed 
in some detailed in the next section.

De Alfaro et al. \cite{c2} noticed a very interesting property of conformal dynamics. Namely, any 
element of the Lie algebra of $sL(2,\mathbb{R})$ can serve as the Hamiltonian provided one changes 
simultaneously the definition of time variable. To see this consider the generator
\\
\begin{equation}
\label{20}
\tilde G = u\tilde H + v\tilde D + \omega \tilde K
\end{equation}
\\
with any real $u$, $v$ and $\omega $. It is a constant of motion. Moreover,
\\
\begin{equation}
\label{21}
\left\{ q,\tilde G \right\} = f(t) \dot q - \frac{1}{2} \dot f(t)q
\end{equation}
where
\begin{equation}
\label{22}
f(t) = u + vt + \omega t^2.
\end{equation}
\\
Define now new time variable and new coordinate
\\
\begin{eqnarray}
\tau &\equiv& \int \frac{dt}{f(t)} = \int \frac{dt}{u + v + \omega t^2}\nonumber\\
Q(\tau ) &\equiv & \frac{q(t)}{\sqrt{f(t)}}.
\label{23}
\end{eqnarray}

Then it is easy to check that
\\
\begin{equation}
\label{24}
\frac{dQ(\tau )}{d\tau } = \left\{ Q(\tau ),\tilde G \right\}.
\end{equation}
\\
Now, eqs. (\ref{23}) define a (time-dependent) point transformation which allow us to compute 
the new Lagrangian
\\
\begin{equation}
L'(Q,\frac{dQ}{d\tau }) = \frac{m}{2}\left( \frac{dQ}{d\tau }\right)^2 - \frac{g}{2Q^2} + \frac{m}{8}\left(v^2 - 4u\omega \right) Q^2.
\label{25}
\end{equation}

Computing the relevant Hamiltonian one finds
\\
\begin{equation}
\label{26}
H' = \tilde G
\end{equation}
\\
as expected. According to the standard Lie-algebraic terminology the generator $\tilde G$ is compact, noncompact or parabolic
depending whether $\Delta \equiv v^2 - 4u\omega $ is negative, positive or zero, respectively.

Let us conclude this short review with the following remark. Eqs. (\ref{23})-(\ref{25}) allow us to generalize the so-called Niederer
transformation \cite{c8} relating free ($g=0$) and harmonic oscillator motions. Indeed, taking $v = 0$, $u = \omega = 1$ we find
that the transformation 
\\
\begin{eqnarray}
\tau &=& \arctan t \nonumber\\
Q(\tau )&=&\frac{q(t)}{\sqrt{1 + t^2}}
\label{27}
\end{eqnarray}
\\
relates the initial Hamiltonian $\tilde H$, eq.(\ref{10}), to the new one
\\
\begin{equation}
\label{28}
\tilde H' = \frac{p^2}{2m} + \frac{g}{2Q^2} + \frac{m}{2} Q^2
\end{equation}
\\
which contains an additional harmonic term.

\section{"Space-time" action of conformal group.}

The conformal algebra (\ref{8}) is isomorphic to $sO(2,1)$ algebra. To see this one defines
\begin{eqnarray}
M^0 &=& - \frac{1}{2} (H + K)\nonumber\\
\label{29}
M^1 &=& -\frac{1}{2}(H-K)\\
M^2 &=& -D\nonumber
\end{eqnarray}
\\
which yields
\begin{equation}
\label{30}
\left[ M^\alpha , M^\beta \right] = - i \epsilon ^{\alpha \beta }\ _{\gamma }M^\gamma , \qquad \alpha ,\beta ,\gamma = 0,1,2
\end{equation}
\\
where $\epsilon ^{012}=\epsilon _{012}=1$ and $g_{\mu \nu }=diag(+--)$.
\\
The basic representation of (\ref{29}) or (\ref{30}) can be taken as
\begin{equation}
\label{31}
\tilde M ^0 = -\frac{1}{2} \sigma _{2}, \qquad \tilde M^1 = - \frac{i}{2}\sigma _1, \qquad \tilde M^2 = -\frac{i}{2}\sigma _3
\end{equation}
\\
or, by virtue of eqs.(\ref{29})
\begin{equation}
H = i\sigma _+, \qquad K = -i\sigma _-, \qquad D = -\frac{i}{2}\sigma _3
\label{32}
\end{equation}
\\
which clearly generates $SL(2,\mathbb{R})$. Due to the isomorphism of algebras $SL(2,\mathbb{R})$ and $SO(2,1)$ are locally isomorphic. 
In fact, the group $SL(2,\mathbb{R})$ is nothing but the group $Spin(2,1)^+$.

$SL(2,\mathbb{R})$ acts nonlinearly on $\mathbb{R}$ (identified with time axis). This can be understood in terms of nonlinear realizations
\cite{c7}. Namely $K$ and $D$ generate a subgroup of $SL(2,\mathbb{R})$ so according to the method of nonlinear realizations one can
define the action of $SL(2,\mathbb{R})$ on $t$-axis by
\\
\begin{equation}
\label{33}
g e^{-itH}= e^{-it'(t,g)H} e^{i\lambda (t,g)D} e^{-ic(t,g)K};
\end{equation}
\\
$t'(t,g)$ is a singular function of their arguments  (eq. (\ref{33}) yields $t' = \frac{t}{(1-ct)}$ in the case of conformal 
transformations). The reason for this singularity is that for $SL(2,\mathbb{R})$, which is semisimple, the exponential parametrization 
does not provide a global map. Therefore, using the method of nonlinear realizations, one should take special care about topological 
subtleties. To this end let us note that $K$ and $D$ generate the Borel subgroup $B$ consisting of the elements of the form \cite{c9}:
\\
 \begin{equation}
 \left(
\begin{array}{cc}
     r & 0 \\
     s  & r^{-1} 
     \end{array}
     \right)
     ,
\qquad r\in \mathbb{R}_+, \quad s \in \mathbb{R}.
\label{34}
\end{equation}
\\
Let $ W = SO(2,\mathbb{R}) \subset SL(2,\mathbb{R})$; any element $g$ of $SL(2,\mathbb{R})$ can be written uniquely as
\\
\begin{equation}
\label{35}
g=wb \quad,\qquad w\in W\quad,\qquad b\in B.
\end{equation}
\\
Therefore, the coset space $SL(2,\mathbb{R} /B)$ can be identified with $W$ (in the "physical" terminology of Ref. \cite{c7} 
this means that the coset manifold is generated by $ H + K $) The standard nonlinear realization is now defined by
\\
\begin{equation}
\label{36}
gw= w'(g,w)b(g,w)
\end{equation}
\\
or $(x^2 + y^2 = 1 = {x'}^2 + {y'}^2)$
\begin{equation}
\label{37}
\left(
\begin{array}{cc}
\alpha &\beta \\
\gamma &\delta 
\end{array}
\right)
\left(
\begin{array}{cc}
x & y\\
-y & x
\end{array}
\right)
= \left(
\begin{array}{cc}
x' & y'\\
-y' & x'
\end{array}
\right)
\left(
\begin{array}{cc}
r & 0\\
s & r^{-1}
\end{array}
\right).
\end{equation}
\\
Eq. (\ref{37}) can be solved uniquely for $x'$, $y'$, $r$ and $s$. However, we need only two relations
\\
\begin{eqnarray}
\alpha y + \beta x &=&r^{-1}y'\nonumber \\
\gamma y + \delta x &=& r^{-1}x'.
\label{38}
\end{eqnarray}
\\
$W$, being topologically a circle can be identified with the projective space $\mathbb{P}^1(\mathbb{R})$; to this end one 
has only to identify the opposite points of the circle. As a result $\mathbb{P}^1(\mathbb{R}) = W/\{I, -I\}$ and $(x,y)$
are homogeneous coordinates; moreover, eqs. (\ref{38}) define the action of $SL(2,\mathbb{R})$ on $\mathbb{P}^1(\mathbb{R})$;
in fact one easily finds that the action of $PSL(2,\mathbb{R}) \equiv  SL(2,\mathbb{R})/\{I,-I\}$ is even faithful.
\\
$\mathbb{P}^1(\mathbb{R})$ can be covered by two maps
\begin{eqnarray}
t=\frac{y}{x} , \qquad x \neq 0\nonumber\\
\tilde t=-\frac{x}{y}, \quad y \neq 0.
\label{39}
\end{eqnarray}
\\
In the common domain $xy \neq 0$ the transition map reads
\begin{equation}
\label{40}
\tilde t = - \frac{1}{t}.
\end{equation}
\\
The minus sign in second equation (\ref{39}) is chosen in order to define $\tilde t$ as increasing function
of $t$ (preservation of time arrow).
\\
The action of $SL(2,\mathbb{R})$ on $\mathbb{P}^1(\mathbb{R})$ is now given by
\begin{eqnarray}
t' &=&\frac{\alpha t + \beta }{\gamma t + \delta }, \qquad \gamma t +\delta \neq 0\nonumber\\
\tilde t' &=& \frac{\delta \tilde t - \gamma }{-\beta \tilde t + \alpha }, \qquad -\beta \tilde t + \alpha \neq 0.
\label{41}
\end{eqnarray}
\\
Within this description $SL(2,\mathbb{R})$ acts regularly on compactified $1+0$ "space-time" manifold $\mathbb{P}^1(\mathbb{R})$ \cite{c10}.
In the common domain of two maps the transition function is given by eq. (\ref{40}). According to the results presented
in the previous section this time redefinition can be viewed as canonical (in fact - point) transformation defined by eqs. (\ref{23})
with $\omega = 1$. The new Hamiltonian, when expressed in old coordinates, coincides with $\tilde K$.

Let us note that the action of $SL(2,\mathbb{R})$ on $\mathbb{P}^1(\mathbb{R})$ defined by eqs. (\ref{41}) implies that the mapping
\begin{equation}
\left(
\begin{array}{cc}
\alpha & \beta \\
\gamma & \delta 
\end{array}
\right)
\longmapsto 
\left(
\begin{array}{cc}
\delta & -\gamma \\
-\beta & \alpha 
\end{array}
\right)
\label{42}
\end{equation}
\\
is an automorphism of $SL(2,\mathbb{R})$. It induces the following automorphism of $sL(2,\mathbb{R})$:
\begin{equation}
\label{43}
D\longmapsto -D, \quad H\longmapsto K, \quad K\longmapsto H
\end{equation}
\\
as expected.

Finally, the following remark is in order. We noted above that the topology of time manifold
is properly taken into account if, in the language of nonlinear realizations, the relevant coset
space is generated by exponentiation of $H + K$. Calling $\theta $ the parameter in front of $H + K$ we find
the following relation
\\
\begin{equation}
\label{44}
t = \tan\theta 
\end{equation}
\\
This again agrees with the findings of Ref.\cite{c2} where the relation (\ref{44}) defines the transformation between
the dynamical evolution parameters corresponding to $H$ and $H + K$. 

\section{The geometry of conformal dynamics.}

The discussion presented in previous sections shows that, at least in the case of repulsive potentials, the $SL(2,\mathbb{R})$
transformations are defined as smooth ones acting on phase space (canonical symmetry transformations) or on coordinate space
(point transformations); in the latter case the nontrivial topology of $1 + 0$-dimensional space-time must be taken into account.\\
The question arises whether the list of $SL(2,\mathbb{R})$ invariant dynamical regular systems is exhausted by the repulsive inverse square
potential; in particular, whether the dynamics corresponding to attractive potential is well-defined. To answer this question, 
let us note that, under the additional assumption that the action of $SL(2,\mathbb{R})$ is transitive, all $SL(2,\mathbb{R})$ invariant 
Hamiltonian systems may be classified using the orbit method \cite{c4}. This has been done in Ref.\cite{c5} and is summarized 
in the present section.\\
The essence of the orbit method consists in the following: given a Lie group G the full list of phase manifolds on which G acts 
transitively as the group of canonical transformations is provided by the set of coadjoint orbits equipped with the Poisson 
structure determined by Kirillov symplectic form.

In the $SL(2,\mathbb{R})$ case the coadjoint action is the same as for $SO(2,1)$ group. Therefore, one has to find the orbits of 
$2+1$-dimensional Lorentz group. The solution is well-known. There are five possible types of orbits ($\xi \equiv (\xi _0,\xi _1,\xi _2)$):
\\
\begin{eqnarray}
\mathcal{H}^+_\sigma &=& \{\xi _\alpha \xi ^\alpha = \sigma ^2,\ \xi _0 > 0 \}\nonumber\\
\mathcal{H}^-_\sigma &=& \{\xi _\alpha \xi ^\alpha = \sigma ^2,\ \xi _0 < 0 \}\nonumber\\
\mathcal{H}_\sigma &=& \{\xi _\alpha \xi ^\alpha = -\sigma ^2 \}
\label{44a}\\
\mathcal{H}^+_0 &=& \{\xi _\alpha \xi ^\alpha = 0,\ \xi _0 > 0 \}\nonumber\\
\mathcal{H}^-_0 &=& \{\xi _\alpha \xi ^\alpha = 0,\ \xi _0 < 0 \}\nonumber
\end{eqnarray}
\\
and the trivial orbit $\xi _0 = 0$. The nondegenerate Poisson structure reads
\\
\begin{equation}
\label{45}
\left\{\xi _\alpha ,\xi _\beta \right\} = -\epsilon _{\alpha \beta }\,^\gamma \xi _\gamma \qquad \left(\epsilon _{012}=1\right).
\end{equation}
\\
The generators of conformal algebra are
\\
\begin{equation}
K = \xi _0 + \xi _1,\qquad H = \xi _0  - \xi _1, \qquad D = \xi _2.
\label{46}
\end{equation}
\\
The canonical equations of motion 
\\
\begin{equation}
\dot \xi _\alpha = \left\{\xi _\alpha , H \right\}
\label{47}
\end{equation}
\\
yield
\\
\begin{eqnarray}
\xi _0(t) &=& \frac{1}{2}\left( \xi _0 - \xi _1 \right) t^2 - \xi _2 t + \xi _0\nonumber\\
\xi _1(t) &=& \frac{1}{2}\left( \xi _0 - \xi _1 \right) t^2 - \xi _2 t + \xi _1
\label{48}\\
\xi _2(t) &=& -\left( \xi _0 - \xi _1 \right) t + \xi _2. \nonumber
\end{eqnarray}
\\

Let us now consider the particular orbits. We start with $\mathcal{H}^+_\sigma$; it is the upper sheet of two-sheeted
hyperboloid. We introduce new variables, $0<x<\infty $, $-\infty<p<\infty$ by the formulae
\\
\begin{eqnarray}
\xi _0 &=& \frac{p^2}{4m} + \frac{\sigma ^2}{mx^2} + \frac{mx^2}{4}\nonumber\\
\label{49}
\xi _1 &=& -\frac{p^2}{4m} - \frac{\sigma ^2}{mx^2} + \frac{mx^2}{4}\\
\xi _2 &=& -\frac{1}{2}xp.\nonumber
\end{eqnarray}
 \\
One easily checks that the eqs. (\ref{49}) provide a smooth map for the orbit $\mathcal{H}^+_\sigma$; moreover, $x$ and $p$ 
are global Darboux coordinates, $\{x,p\}=1$. The Hamiltonian $H = \xi _0  - \xi _1$ coincides with the standard 
Hamiltonian with repulsive potential.\\
As far as $\mathcal{H}^-_\sigma$ is concerned note that $\xi _0 \rightarrow -\xi _0$, $\xi _1 \rightarrow -\xi _1$, 
$\xi _2 \rightarrow -\xi _2$, is an automorphism of Poisson structure. Therefore, one obtains the relevant modification
of the mapping (\ref{49}) which leads to negative definite Hamiltonian
\\
\begin{equation}
\label{50}
H = -\frac{p^2}{2m} - \frac{2\sigma ^2}{mx^2}.
\end{equation}

The case of $\mathcal{H}_\sigma $ is much more interesting. The one-sheeted hyperboloid provide the manifold which cannot
be covered by one map. It is not difficult to construct two maps covering the whole manifold and defining local Darboux coordinates \cite{c5}.
However, it is not necessary. Instead, we define two maps which almost cover the hyperboloid. To this end we consider the intersection of
the hyperboloid with the plane $\xi _0 + \xi _1 = 0$. This cuts the manifolds into two submanifolds (see Fig.1)

\begin{figure}[h!]
\includegraphics{obrazki-4.eps}
\caption{}
\end{figure}
 
\begin{equation}
\label{51}
M_{\pm } = \left\{ \xi ^\alpha  \xi _\alpha = -\sigma ^2 \mid \xi _0 + \xi _1 \ ^<_> 0 \right\}
\end{equation}
\\
Together with two lines $\xi _0 + \xi _1 = 0$, $\xi _2 = \pm \sigma $ they cover whole hyperboloid. We parametrize $M_+$ and $M_-$ as follows\\
$M_+$:
\begin{eqnarray}
\xi_0 &=& \frac{p^2}{4m} - \frac{\sigma ^2}{mx^2} + \frac{mx^2}{4}, \qquad 0 < x < \infty,\quad -\infty < p < \infty \nonumber\\
\xi_1 &=& -\frac{p^2}{4m} + \frac{\sigma ^2}{mx^2} + \frac{mx^2}{4}
\label{52}\\
\xi_2 &=& -\frac{1}{2}xp\nonumber
\end{eqnarray}
$M_-$:
\begin{eqnarray}
\xi_0 &=& -\frac{p^2}{4m} + \frac{\sigma ^2}{mx^2} - \frac{mx^2}{4}, \qquad -\infty < x < 0,\quad -\infty < p < \infty \nonumber\\
\xi_1 &=& \frac{p^2}{4m} - \frac{\sigma ^2}{mx^2} - \frac{mx^2}{4}
\label{53}\\
\xi_2 &=& -\frac{1}{2}xp.\nonumber
\end{eqnarray}
\\
Let us find the trajectories in terms of old and new variables. To this end we find from eqs. (\ref{48})
\\
\begin{equation}
\label{54}
\xi _0(t) + \xi _1(t) = \left( \xi _0 - \xi _1 \right) t^2 - 2\xi _2 t + \left( \xi _0 + \xi _1\right).
\end{equation}
\\
The equation $\xi _0(t) + \xi _1(t) = 0$ has two solutions for $\xi _0 - \xi _1 \neq 0$ and one for $\xi _0 - \xi _1 = 0$.
So the solution traverses $M_+$ or $M_-$ in finite time depending on whether the energy $E = \xi _0 - \xi _1$ is negative
or positive, respectively; the only exception is zero-energy motion. Taking into account that the allowed $x$ region is
$E\geq -\frac{2\sigma ^2}{mx^2}$ for $x>0$ and $E<\frac{2\sigma ^2}{mx^2}$ for $x<0$ we arrive at the picture of motion
in terms of $x$ variable which is illustrated on Fig.2.

\begin{figure}[h!]
\includegraphics{obrazki-2.eps}
\caption{}
\end{figure}

 We conclude that the description of dynamics in terms of positive 
values of $x$ coordinate is incomplete. The singularity related to the effect of "falling on the center" in finite time
is spurious as an artifact of the choice of coordinates in symplectic manifold.

Finally, consider the case of light cone. The forward $(\xi _0 > 0)$ and backward $(\xi_0 < 0)$ cones are separated by one 
point $\xi = 0$ which by itself forms an orbit. The intersection of both cones with the plane $\xi_0 + \xi_1 = 0$ forms now 
single straight line. Eq. (\ref{54}) can be rewritten as 
\\
\begin{equation}
\xi_0(t) + \xi_1(t) = \left( \xi_0 - \xi_1\right )\left( t - \frac{\xi_2}{\xi_0 -\xi_1}\right) ^2 .
\label{55}
\end{equation}
\\
So $\xi_0(t) + \xi_1(t) \geq 0 \ (\xi_0(t) + \xi_1(t) \leq 0)$ for forward (backward) cone. The points $\xi_0 = \xi_1$, $\xi_2 = 0$
are fixed points of dynamics. According to eq. (\ref{55}) any other trajectory crosses the line $\xi_0 + \xi_1 = 0$ exactly once.
This is depicted on Fig.3. 

\begin{figure}[h!]
\includegraphics{obrazki-3.eps}
\caption{}
\end{figure}

Now, we introduce the $x-p$ variables. Consider the forward cone. It can be parametrized by $\xi_1$ and $\xi_2$ variables
running over the plane with origin deleted. Define the canonical variables $x,p$ by
\\
\begin{eqnarray}
\xi_1 &=& -\frac{p^2}{4m} + \frac{mx^2}{4} \qquad -\infty < x < \infty \ ,\quad -\infty < p < \infty\nonumber\\
\xi_2 &=& -\frac{1}{2}xp.
\label{56}
\end{eqnarray}
\\
Then $H = \xi_0 - \xi_1$ gives 
\\
\begin{equation}
\label{57}
H = \frac{p^2}{2m}
\end{equation}
\\
which yield free motion. Let us analyze in more detailed the mapping (\ref{56}). Introducing the complex variables
\\
\begin{eqnarray}
\xi &=& \xi_1 + i\xi_2\nonumber\\
u &=& \frac{\sqrt{m}}{2}x - \frac{ip}{2\sqrt{m}}
\label{58}
\end{eqnarray}
\\
eqs.( \ref{56}) can be rewritten as
\\
\begin{equation}
\label{59}
\xi = u^2 .
\end{equation}
\\
Consider $\xi$ plane with the origin deleted and the cut along negative $\xi_1$-axis; note that the cut corresponds to
the line $\xi_0 + \xi_1 = 0$. Now, $\xi \mapsto u$ is analytic on two-sheeted Riemann surface. Knowing $\xi(t)$ one easily 
finds the trajectory $u = u(t)$. Typical trajectory (with $\xi_0 - \xi_1 > 0$) is drawn on Fig.4.

\begin{figure}[h!]
\includegraphics{obrazki-1.eps}
\caption{}
\end{figure}

 Equation of motion (\ref{48})
imply
\\
\begin{equation}
\label{60}
\xi = \left( \sqrt{\frac{E}{2}}\left( t - \frac{\xi_2}{\xi_0 - \xi_1}\right) - i\sqrt{\frac{E}{2}}\right)^2.
\end{equation}
\\

Let us assume we start from the upper sheet. By comparing eqs. (\ref{58})-(\ref{60}) one finds
\\
\begin{eqnarray}
x &=& - \sqrt{\frac{2E}{m}}\left(t - \frac{\xi_2}{\xi_0 - \xi_1}\right)\nonumber\\
p &=& -\sqrt{2mE}.
\label{61}
\end{eqnarray}
\\
Similarly, starting from the lower sheet one finds solutions with the minus sign on the right-hand side replaced by plus.

\section{The action of conformal $SL(2,\mathbb{R})$ group.}

We have shown in the previous section that the most general Hamiltonian mechanics with conformal $SL(2,\mathbb{R})$ invariance group
(which acts transitively on phase space) is the standard conformal mechanics both with repulsive as well as attractive potential
(together with free motion corresponding to vanishing potential). In the latter case the standard description is not complete
because it does not take into account nontrivial topology of phase manifold. In particular, the "falling on the center" phenomenon
is an artifact of the choice of local canonical variables.

Let us now consider the action of conformal group $SL(2,\mathbb{R})$ within the global formulation presented above. The symmetry generators read
\\
\begin{eqnarray}
H(t) &=& H\nonumber\\
D(t) &=& D + tH
\label{62}\\
K(t) &=& K + 2tD + t^2H .\nonumber
\end{eqnarray}

In what follows we consider only the most interesting case of special conformal transformations generated by $K(t)$. 
By virtue of last eq. (\ref{62}) we find
\\
\begin{equation}
\label{63}
K(t) = \left( 1 + t^2 \right) \xi_0 + \left( 1 - t^2\right) \xi_1 + 2t\xi_2 .
\end{equation}
\\
Let $c$ be the conformal parameter. The conformal transformations are defined by the equation
\\
\begin{equation}
\label{64}
\frac{\partial \xi_\alpha (t,c)}{\partial c} = \left\{ K(t), \xi _\alpha (t,c)\right\}
\end{equation}
\\
which, via (\ref{63}), yields
\\
\begin{eqnarray}
\frac{\partial \xi_0}{\partial c} &=& -\left( 1 - t^2\right)\xi_2 + 2t\xi_1\nonumber\\
\label{65}
\frac{\partial \xi_1}{\partial c} &=& \left( 1 + t^2\right)\xi_2 + 2t\xi_0\\
\frac{\partial \xi_2}{\partial c} &=& -\left(1+t^2\right)\xi_1 - \left(1 - t^2\right)\xi_0 .\nonumber
\end{eqnarray}
\\
This set of equations is easily solvable and gives:
\\
\begin{eqnarray}
\xi_0(t,c) &=& \xi_0(t,0) +c\left( 2t\xi_1(t,0) + \left( t^2 - 1 \right) \xi_2(t,0) \right) +\nonumber\\
 & & 
+\frac{c^2}{2}\left( \left( t^2 +1\right)^2\xi_0(t,0) + \left( 1 - t^4\right)\xi_1(t,0) + 2t \left(t^2 + 1\right)\xi_2(t,0)\right)\nonumber\\
\label{66}
\xi_1(t,c) &=& \xi_1(t,0) + c\left( 2t\xi_0(t,0) + \left( t^2 + 1 \right)\xi_2(t,0)\right)+\\
& &
+ \frac{c^2}{2}\left(\left(t^4 - 1\right)\xi_0(t,0) - \left(t^2 - 1\right)^2\xi_1(t,0) + 2t\left(t^2 - 1\right)\xi_2(t,0)\right)\nonumber\\
\xi_2(t,c) &=& \xi_2(t,0) + c\left(\left(t^2 - 1\right)\xi_0(t,0) - \left( t^2 + 1\right)\xi_1(t,0)\right)+\nonumber\\
& & 
+\frac{c^2}{2}\left(-2t\left(t^2 + 1\right)\xi_0(t,0) + 2t\left( t^2 -1\right)\xi_1(t,0) - 4t^2\xi_2(t,0)\right) .\nonumber
\end{eqnarray} 

We see that the transformations are perfectly regular. This is also the case for the remaining transformations (time translations
and dilatations) and, consequently, for the whole $SL(2,\mathbb{R})$ group.

Consider now the above conformal transformations expressed in terms of $x$ and $p$ variables. Let us take first $\mathcal H ^+_\sigma $. 
One can easily find that eqs. (\ref{66}) can be rewritten as
\\
\begin{eqnarray}
x(c) &=& \sqrt{\left( x(1 + ct) - \frac{pct^2}{m}\right)^2 + \frac{4\sigma ^2c^2t^4}{m^2x^2}}\nonumber\\
x(c)p(c) &=& xp\left(1 - 2c^2t^2\right) + mc\left(1 + ct\right) x^2 - 2ct^2\left( 1-ct\right)H .
\label{67}
\end{eqnarray}
\\
We see that this transformation is perfectly regular in the halfplane: $x>0$, $p\in \mathbb R$. The same conclusion can be drawn for 
$\mathcal H^-_\sigma $. \\
For  $\mathcal H_\sigma $ the situation is more involved. Assume we start from the point $\xi \in M_+$, i.e. $\xi_0 + \xi_1 > 0$.
The form of transformation depends on whether $\xi_0(c) + \xi_1(c) \ ^>_< 0$ i.e. whether the image of initial point lies in $M_+$ 
or $M_-$. A simple computation based on eqs. (\ref{52}) and (\ref{53}) gives
\\
\begin{equation}
\label{68}
x(c)=
\left\{
\begin{array}{ll}
\sqrt{\left( x(1 + ct) - \frac{pct^2}{m}\right)^2 - \frac{4\sigma ^2 c^2 t^4}{m^2 x^2}}, &
\textrm{for $\left( x(1 + ct) - \frac{pct^2}{m}\right)^2 - \frac{4\sigma ^2 c^2 t^4}{m^2 x^2} >0$}\\
\\
-\sqrt{ \frac{4\sigma ^2 c^2 t^4}{m^2 x^2}-\left(x(1 + ct) - \frac{pct^2}{m}\right)^2}, &
\textrm{for  $\left( x(1 + ct) - \frac{pct^2}{m}\right)^2 - \frac{4\sigma ^2 c^2 t^4}{m^2 x^2} <0$}
\end{array}
\right.
\end{equation}
\\
while the second eq. (\ref{67}) remains unchanged. Our transformation is undefined if 
$\left( x(1 + ct) - \frac{pct^2}{m}\right)^2 - \frac{4\sigma ^2 c^2 t^4}{m^2 x^2} = 0$. 
The reason is that the image lies then on one borderlines between $M_+$ and $M_-$. This singularity is again
an artifact of the choice of canonical coordinates.

Finally, consider $\mathcal H^+_0 $. We find the following transformation properties:
\\
\begin{eqnarray}
x(c) &=& x(1+ct) - \frac{pct^2}{m}\nonumber\\
p(c) &=& p(1-ct) +mcx
\label{69}
\end{eqnarray}
\\
which are regular everywhere on $(x,p)$-plane. The same holds for $\mathcal H_0 ^-$.

Consider now the transformations which comprise the change of time variable.
\\
\begin{equation}
\label{70}
t' = \frac{t}{1-ct} .
\end{equation}
\\
It is described by the "reduced" generators which do not contain the Hamiltonian while the time variable
$t$ is replaced by $t'$.\\
The equations defining such transformations read
\\
\begin{eqnarray}
\frac{d\xi _\alpha (t'(t,c),c)}{dc} &=& \left\{ K_r,\xi _\alpha (t'(t,c),c)\right\}\nonumber\\
K_r(t,c) &=& \xi _0(t'(t,c),c) + \xi _1(t'(t,c),c) + \frac{2t}{1-ct}\xi _2(t'(t,c),c)
\label{71}
\end{eqnarray}
\\
or, explicitly
\\
\begin{eqnarray}
\frac{d\xi _0}{d c} &=& -\xi_2 + \frac{2t}{1-ct}\xi_1\nonumber\\
\frac{d\xi _1}{d c} &=& \xi_2 + \frac{2t}{1-ct}\xi_0
\label{72}\\
\frac{d\xi _2}{d c} &=& -\left( \xi_0 + \xi_1\right) .\nonumber
\end{eqnarray}
\\
Eqs.(\ref{72}) can be easily integrated to give the finite form of transformations
\\
\begin{eqnarray}
\xi_0(t',c) + \xi_1(t',c) &=& \frac{\xi_0(t,0) + \xi_1(t,0)}{(1-ct)^2}\nonumber\\
\xi_0(t',c) - \xi_1(t',c) &=& \left( \xi_0(t,0) - \xi_1(t,0)\right)(1-ct)^2 + \nonumber\\
 & &
\left(\xi_0(t,0) + \xi_1(t,0)\right)c^2 - 2c(1-ct)\xi_2(t,0)
\label{73}\\
\xi_2(t',c) &=& \xi_2(t,0) - \frac{c}{1-ct}\left(\xi_0(t,0) +\xi_1(t,0)\right) .\nonumber
\end{eqnarray}
\\
Using eq. (\ref{70}) and the equations of motion (\ref{48}) one easily verifies that eqs.(\ref{66}) and (\ref{73}) are equivalent.

Let us again describe eqs. (\ref{73}) from the point of view of $x,p$ coordinates. In the case of $\mathcal H^+_\sigma $ one finds
\\
\begin{eqnarray}
x'(t') &=& \frac{x(t)}{1-ct}\nonumber\\
p'(t') &=& (1-ct)p(t) + mcx(t) .
\label{74}
\end{eqnarray}
\\
So, for $c$ small enough, $ct<1$, we get a regular mapping (the same conclusion holds true for $\mathcal H^-_\sigma $). In the case
of $\mathcal H_\sigma $ we obtain the mappings of $M_+$ (respectively $M_-$) onto $M_+$ (respectively $M_-$).

As explained in Sec.III the singularity in conformal time transformations are related to $0+1$ space-time nontrivial topology. 
To get rid of it we change the map according to eq. (\ref{40}). This induces the automorphism (\ref{43}). In order to find 
the relevant transformation on coadjoint orbit we rewrite eqs. ({\ref{43}) in terms of $\xi$ variables (here $\tilde \xi$ are new $\xi$ 
variables:
\\
\begin{eqnarray}
&&\xi_0 - \xi_1 = \tilde \xi_0 + \tilde \xi_1 + 2\tilde t \tilde \xi_2 + \tilde t^2\left(\tilde \xi_0 - \tilde \xi_1\right)\nonumber\\
&&\xi_0 + \xi_ 1 + 2t\xi_2 + t^2\left( \xi_0 - \xi_1\right) = \tilde \xi_0 - \tilde \xi_1
\label{75}\\
&&-\xi_2 - t\left(\xi_0 - \xi_1\right) = \tilde \xi_2 + \tilde t\left( \tilde \xi_0 - \tilde \xi_1\right)\nonumber
\end{eqnarray}
\\
or
\\
\begin{eqnarray}
\tilde \xi_0 + \tilde \xi_1 &=& \frac{\xi_0 + \xi_1}{t^2}\nonumber\\
\tilde \xi_0 - \tilde \xi_1 &=& \xi_0 + \xi_1 + 2t\xi_2 + t^2\left( \xi_0 - \xi_1 \right)
\label{76}\\
\tilde \xi_2 &=& \xi_2 + \frac{\xi_0 + \xi_1}{t} .\nonumber
\end{eqnarray}
\\
Eq. (\ref{76}) defines an automorphism (it preserves the Poisson structure) of coadjoint orbit. Moreover, one easily verifies that
the dynamical equations (\ref{47}) are invariant under (\ref{40}) and (\ref{76}). The action of conformal transformations on new 
time variable $\tilde t$ reads
\\
\begin{equation}
\label{77}
\tilde t' = \tilde t + c
\end{equation}
as expected.

\section{Time measurement}

Up to now the time variable $t$ played the role of abstract dynamical parameter. However, in any physically reasonable theory 
the method of relating this parameter to physically measurable quantities should be indicated. To this end one has to construct 
a "clock", i.e. physically realizable device which can serve to quantify the time flow.\\
Our clock is defined as $SL(2,\mathbb{R})$-invariant dynamical system based on $\mathcal H^+_0$ orbit. Let us rewrite last eq. (\ref{48})
in the form
\\
\begin{equation}
\label{78}
\frac{\xi_2(t)}{\xi_1(t) - \xi_0(t)} - \frac{\xi_2(t_0)}{\xi_1(t_0) - \xi_0(t_0)} = t - t_0 .
\end{equation}
\\
We further define the "pointer" of our clock as the dynamical variable $\frac{\xi_2}{\xi_1 - \xi_0}$. Then time is simply given 
by the value of this variable
\\
\begin{equation}
\label{79}
t = \frac{\xi_2}{\xi_1 - \xi_0} \equiv -\frac{\xi_0 +\xi_1}{\xi_2}
\end{equation}
\\
This equation stressed the projective nature of time variable. Moreover, it agrees with the definition of space-time coordinates
which results from considering the space-time action of conformal group as reduction of its action on higher dimensional 
light cone \cite{c6}.

Using eqs. (\ref{73}) one confirms that the definition of time variable provided by eq. (\ref{79}) leads to the proper transformation rule
\\
\begin{equation}
\label{80}
t' = \frac{t}{1 - ct}
\end{equation}
\\
as expected.

\section{Conclusions}

Let us conclude the discussion of the preceding sections. All $SL(2,\mathbb{R})$ invariant Hamiltonian systems with transitive action 
of conformal group are classified with the help of orbit method. The admissible phase manifolds are one- and two-sheeted hyperboloids
and forward or backward cones. Both the dynamics and the group action are given by the smooth maps of phase manifolds. 
The singularities appear only on the level of Darboux coordinates. They can be avoided by careful inspection of group action in terms of 
local maps. 

In the case of two-sheeted hyperboloid  each sheet can be covered by one Darboux map. One can then pass to the global Lagrangian
formalism and ask if the conformal transformations may be viewed as point transformations. This appears to be possible provided the time
variable transforms nontrivially. The resulting transformation rule is singular. The singularity can be again removed by considering
the time variable as parameterizing nontrivial projective manifold.\\
Two alternative forms of conformal transformations (with and without affecting the time variable) are possible also in topologically nontrivial
case of one-sheeted hyperboloid.

Once the role of time variable as parameterizing (locally) the projective manifold is recognized both definitions of the action of conformal 
$SL(2,\mathbb{R})$ group yield smooth transformations. It is interesting to note that the geometrical picture presented provides a natural framework
for the flexibility in the choice of time variable described in the paper by Alfaro et al. \cite{c2}

Finally, we would like to stress that the formalism presented here allows to deal with repulsive and attractive potentials; the "falling on the center"
phenomenon which appears in the latter case is an artifact of the choice of local Darboux coordinates.

\section{Acknowledgments}

I would like to thank Professors Krzysztof Andrzejewski, Piotr Kosiński and Paweł Maślanka for helpful discussions and useful remarks.\\
Dr. Bartosz Zieliński is acknowledged for helping in preparing figures.\\
This work is supported in part by MNiSzW Grant No.N202331139.

\end{document}